\pdfoutput=1

\documentclass[11pt]{article}

\usepackage[final]{acl}

\usepackage{times}
\usepackage{latexsym}

\usepackage[T1]{fontenc}

\usepackage[utf8]{inputenc}

\usepackage{microtype}

\usepackage{inconsolata}

\usepackage{graphicx}
\usepackage{listings}
\usepackage{xcolor}

\usepackage{amssymb}
\usepackage{tikz}
\usepackage{hyperref}

\newcommand{\RadioButton}{%
    \tikz[baseline=-0.5ex]{
        \draw[thick] (0,0) circle (0.15); 
        \fill[white] (0,0) circle (0.14); 
    }
}

\renewcommand{\CheckBox}{$\square$\hspace{.5em}} 

\usepackage{fontawesome}

\lstdefinelanguage{json}{
    basicstyle=\ttfamily\tiny,  
    numbers=left,                
    numberstyle=\scriptsize\color{gray}, 
    stepnumber=1,                
    showstringspaces=false,      
    breaklines=true,             
    frame=lines,                 
    literate=
     *{0}{{{\color{blue}0}}}{1}%
      {1}{{{\color{blue}1}}}{1}%
      {2}{{{\color{blue}2}}}{1}%
      {3}{{{\color{blue}3}}}{1}%
      {4}{{{\color{blue}4}}}{1}%
      {5}{{{\color{blue}5}}}{1}%
      {6}{{{\color{blue}6}}}{1}%
      {7}{{{\color{blue}7}}}{1}%
      {8}{{{\color{blue}8}}}{1}%
      {9}{{{\color{blue}9}}}{1}%
      {:}{{{\color{red}{:}}}}{1}%
      {,}{{{\color{red}{,}}}}{1}%
      {\{}{{{\color{gray}{\{}}}}{1}%
      {\}}{{{\color{gray}{\}}}}}{1}%
      {[}{{{\color{gray}{[}}}}{1}%
      {]}{{{\color{gray}{]}}}}{1},
    keywordstyle=\color{orange}\bfseries,   
    stringstyle=\color{green},              
    morekeywords={:}, 
    tabsize=2
}

\title{DataLens: Enhancing Dataset Discovery via Network Topologies}

\author{Anaïs Ollagnier \and Aline Menin \\
Université Côte d’Azur, CNRS, Inria, I3S\\
Sophia Antipolis, France \\
  \texttt{\{anais.ollagnier,aline.menin\}@inria.fr}}

\begin{document}
\maketitle
\begin{abstract}
The rapid growth of publicly available textual resources, such as lexicons and domain-specific corpora, presents challenges in efficiently identifying relevant resources. While repositories are emerging, they often lack advanced search and exploration features. Most search methods rely on keyword queries and metadata filtering, which require prior knowledge and fail to reveal connections between resources. To address this, we present DataLens, a web-based platform that combines faceted search with advanced visualization techniques to enhance resource discovery. DataLens offers network-based visualizations, where the network structure can be adapted to suit the specific analysis task. It also supports a chained views approach, enabling users to explore data from multiple perspectives. A formative user study involving six data practitioners revealed that users highly value visualization tools—especially network-based exploration—and offered insights to help refine our approach to better support dataset search.
\end{abstract}

\section{Introduction}

Advances in artificial intelligence and the open data initiative have promote a exponential growth of publicly available datasets. 
Numerous efforts have been made to improve dataset search, navigation, and discovery~\cite{DBLP:conf/dasfaa/MaierMT14,DBLP:journals/vldb/ChapmanSKKIKG20,10.1145/3626521}. Recent studies have treated this task as an information retrieval problem, improving query handling, data management, and ranking~\cite{DBLP:journals/jiis/DimitrakisST20,DBLP:journals/jdiq/MountantonakisT20,DBLP:journals/pvldb/CasteloRSBCF21}. Despite ongoing efforts to improve findability, accessibility, interoperability, and reusability (e.g., \textit{RDA Data Discovery Paradigm Interest Group}~\footnote{\url{https://www.rd-alliance.org/groups/data-discovery-paradigms-ig}}, \textit{FAIRsharing}\footnote{\url{https://fairsharing.org/}}), significant challenges remain. 
%
%
Metadata quality is a major challenge when searching for machine learning (ML) resources~\cite{DBLP:journals/vldb/ChapmanSKKIKG20,DBLP:journals/patterns/PaulladaRBDH21}. Despite the advances of ML~\cite{10.1007/978-3-031-08751-6_30}, web semantic~\cite{10.1007/978-3-030-33220-4_23}, and metadata aggregation techniques to generate, structure, and enrich metadata, the uncertainty of generated metadata and 
general lack of standardization limit the potential of aggregation techniques as they hinder data integration, reduce search accuracy, and obscure meaningful relationships between datasets. Combined with traditional search methods that rely on keywords and filters to refine results, the sheer volume of information often makes exploration challenging, as suggested resources remain too numerous and obscure potential related datasets. Data visualization is well-known to aid human reasoning by using interactive tools that visually highlight and reveal these connections. Despite its wide use to support visual analysis of individual datasets or data points, their usage to support dataset discovery remains unexplored~\cite{DBLP:journals/cg/LiuBV14}. 

In this paper, we propose \emph{Datalens}\footnote{\textit{DataLens} is available at \url{https://dataviz.i3s.unice.fr/datalens/}}, a dataset search tool that integrates network-based visualizations and chained views to support the discovery of ML resources. Similar to existing approaches, we employ faceted search techniques to deal with large volumes of data. Then, using networks, our approach uncovers hidden relationships between ML resources, such as commonly supported ML tasks and models. Additionally, with chained views, the tool allow the user to narrow the exploration through multiple interconnected visualization techniques, each presenting a different perspective to the data. 
We demonstrate our approach through a set of use case scenarios and a formative evaluation involving 6 data practitioners. 


\section{\textit{DataLens} Overview}

The \emph{DataLens} approach integrates three key components: (i) a network customization panel that allows users to configure the network topology based on relationships pertinent to their search, (ii) a filtering panel that enables users to refine their focus on relevant datasets, and (iii) a visualization tool (MGExplorer~\cite{menin2021mgexp}) that facilitates multi-perspective exploration of ML resources.

\subsection{The Data}

In this paper, we explore data from the Hugging Face catalogue\footnote{\url{https://github.com/huggingface/datasets}}~\cite{DBLP:conf/emnlp/LhoestMJTPPCDPT21}, which includes over 212,000 datasets in more than 8,000 languages, sourced from their API\footnote{\url{https://huggingface.co/docs/hub/api}}. Each data record features 
descriptive information, such as the dataset's \texttt{modality} (indicating the type of data, e.g. text, audio, video), \texttt{task category} (defining the broad types of ML tasks the dataset supports), and \texttt{language}. One may need to find data for training models, or conversely, to identify models trained on specific datasets. Thus, we augmented the datasets' metadata with pre-trained models information, available on Hugging Face (see Listing~\ref{lst:datasetcard}). 

\begin{figure}[!ht]
 \begin{lstlisting}[language=json, caption=Example dataset metadata., numbers=none, label=lst:datasetcard, xleftmargin=0pt, xrightmargin=0pt]
{ "id": "amirveyseh/acronym_identification",
"author": "amirveyseh",
"created_at": "2022-03-02T23:29:22+00:00",
"last_modified": "2024-01-09T11:39:57+00:00",
"downloads": 154,
"likes": 19,
"paperswithcode_id": "acronym-identification",
"tags": [
	"task_categories:token-classification",
	"annotations_creators:expert-generated",
	"language_creators:found",
	"multilinguality:monolingual",
	"source_datasets:original",
	"language:en",
	"license:mit",
	"size_categories:10K<n<100K",
	"format:parquet",
	"modality:text",
	"library:datasets",
	"library:pandas",
	"library:mlcroissant",
	"library:polars",
	"model:deberta-v3-base-tasksource-nli",
	"model:deberta-v3-large-tasksource-nli",
	"model:deberta-v3-xsmall-tasksource-nli",
], ... }  ,
"description": "[...]" }
	\end{lstlisting}
\end{figure}

\subsection{Interactive Customization of Network}

\begin{figure}[!ht]
    \centering
    \includegraphics[width=.9\linewidth]{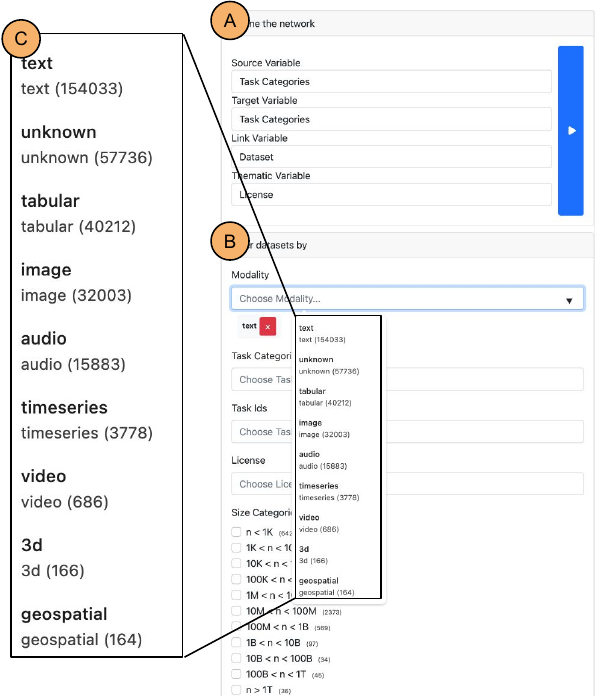}
    \caption{Network customization and faceted search. (A) a panel for customizing the graph topology, (B) a filtering panel for refining the search to focus on relevant datasets. Each search field in the filtering panel includes a list of potential values along with the corresponding number of associated datasets (C).}
    \label{fig:dashboard_Datalens}
\end{figure}

 \begin{figure*}[!ht]
 	\centering
 	\includegraphics[width=\linewidth]{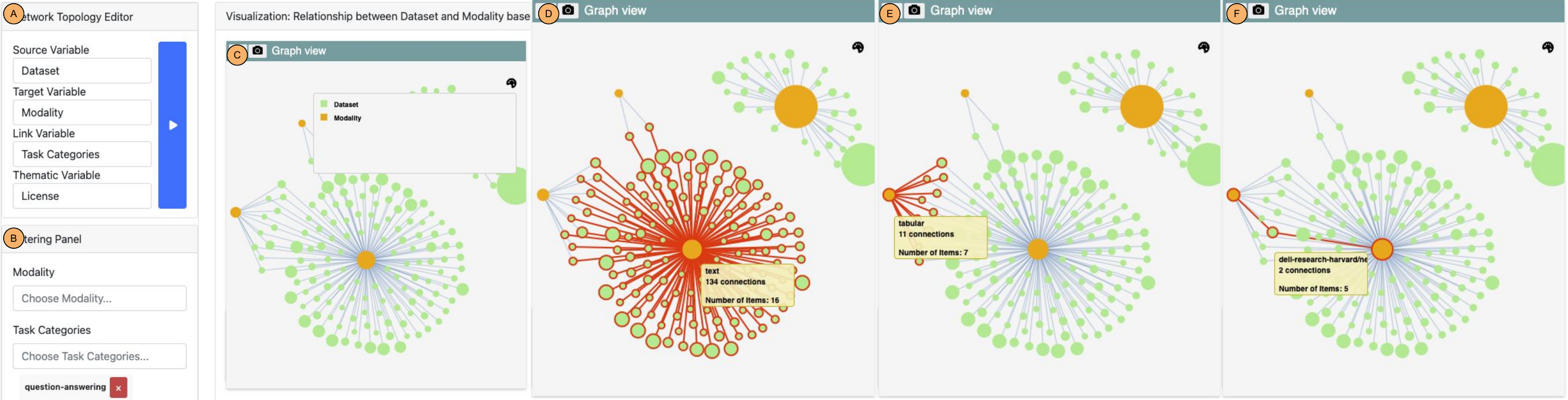}
 	\caption{In Use Case Scenario~\ref{ssec:scenario1}: (A) \textit{Dataset} serve as source variable and \textit{Modality} as the target, linked together by shared \textit{Task Categories}. (B) Data is filtered by selecting \textit{question-answering} (QA) and dataset size as \textit{1M<n<10M}. (C) The graph view depicts datasets and modalities as light-green and orange nodes, respectively. Node size reflects the count of associated tasks. (D-F) Hovering over the circles reveal detailed information such as the number of connections (datasets/modalities) and unique items (links/tasks).}
 	\label{fig:scenario1}
 \end{figure*}

Users can discover meaningful relationships through the \emph{Network Topology Editor} (Fig.~\ref{fig:dashboard_Datalens}A). For instance, users can define nodes as \texttt{task\_categories} interconnected by the \texttt{datasets} that support them, or configure nodes as \texttt{datasets} linked by shared \texttt{license}. The editor facilitates the task through interactive \textit{combobox} menus, enabling users to select the variables that define the \texttt{source} and \texttt{target} nodes, the \texttt{link} between nodes, and the \texttt{thematic} information that describes the connections between nodes. For example, in a network where the nodes represent \texttt{task\_categories} and the links represent \texttt{datasets}, thematic information characterizes the latter through aspects such as \texttt{license}, \texttt{language}, \texttt{format}, \texttt{modality}, and \texttt{size}. 

The \emph{Filtering Panel} (Fig.~\ref{fig:dashboard_Datalens}B) helps users to focus on interesting information. 
For example, if a user is interested in exploring datasets that address at least a given ML task, such as \emph{audio classification}, they can select this option from the combobox menus. The resulting network will then display only those datasets that support at least this task, while also illustrating the relationships with other tasks covered by the filtered datasets. In a more exploratory setting, where the user does not begin with a specific hypothesis, we assist users in discovering potentially relevant resources by displaying the count of datasets associated next to each value in the comboboxes (Fig.~\ref{fig:dashboard_Datalens}D). This allows users to get an overview of the significance of each feature, based on how frequently it appears across datasets.

\subsection{Multi-perspective Exploration}

Our method builds upon MGExplorer's visualization capabilities~\cite{menin2021mgexp}. It facilitates data exploration using a variety of interconnected visualization techniques. These techniques include graph, pairwise relationship, temporal distribution, and listing views. Each view is interactively generated during exploration. The user right-click on an element of interest (e.g., a node in the graph view) and, through a contextual menu, selects and launches a new view to further explore the data. Each new view is then created using a filtered subset of the data, for example, only displaying information related to the selected node in the graph view. Relevant views will be presented and explained along with the use case scenarios hereafter.

\section{Use Case Scenarios}


\subsection{Exploring Multimodal Datasets for Performing the Question-Answering Task}
\label{ssec:scenario1}

Alice is developing a new question-answering (QA) model for a chatbot application. To find relevant datasets for training, she aims to explore datasets that support the QA task while also considering their use across different modalities (e.g., text and tabular) to expand her model’s applicability. She begins her exploration by defining a network topology that aligns with her task. She sets \texttt{dataset} and \texttt{modality} as nodes (i.e. source and target variables, respectively), with \texttt{task categories} as links (Fig.\ref{fig:scenario1}A). Since her focus is on QA, she filters the data by selecting \texttt{question-answering} under \emph{Task Categories} and further refines her selection by choosing datasets with a size between 1M and 10M in the filtering panel (Fig.\ref{fig:scenario1}B). This configuration enables her to visualize the relationship between datasets and data modalities based on shared tasks, including at least QA.
 
Alice starts the visualization by clicking \emph{play}, generating a bipartite graph (Fig.~\ref{fig:scenario1}C) that links datasets (light green) to the modalities (orange) they support for QA or other related ML tasks. Using the hover feature, she explores the orange nodes to examine the associated modalities. She observes that the \texttt{text} modality has the highest number of connections in the network (Fig.~\ref{fig:scenario1}D). The tooltip reveals that \texttt{text} is supported by 134 datasets (i.e., connected nodes) through 16 items (tasks), with 15 potentially related to \texttt{QA}. Additionally, she identifies a shared modality among these datasets. As illustrated in Fig.~\ref{fig:scenario1}E, the \texttt{tabular} modality is linked to 11 datasets, all of which also connect to the \texttt{text} modality. The tooltip indicates that these datasets span 7 different tasks, supported in both \texttt{text} and \texttt{tabular} modalities. One example of a dataset facilitating QA-related tasks across both modalities is \texttt{dell-research-harvard/newswire} (Fig.~\ref{fig:scenario1}F).



\subsection{Expanding QA Applications by Exploring Datasets for Multiple Tasks}
\label{ssec:scenario2}

\begin{figure*}[!ht]
    \centering
    \includegraphics[width=.9\linewidth]{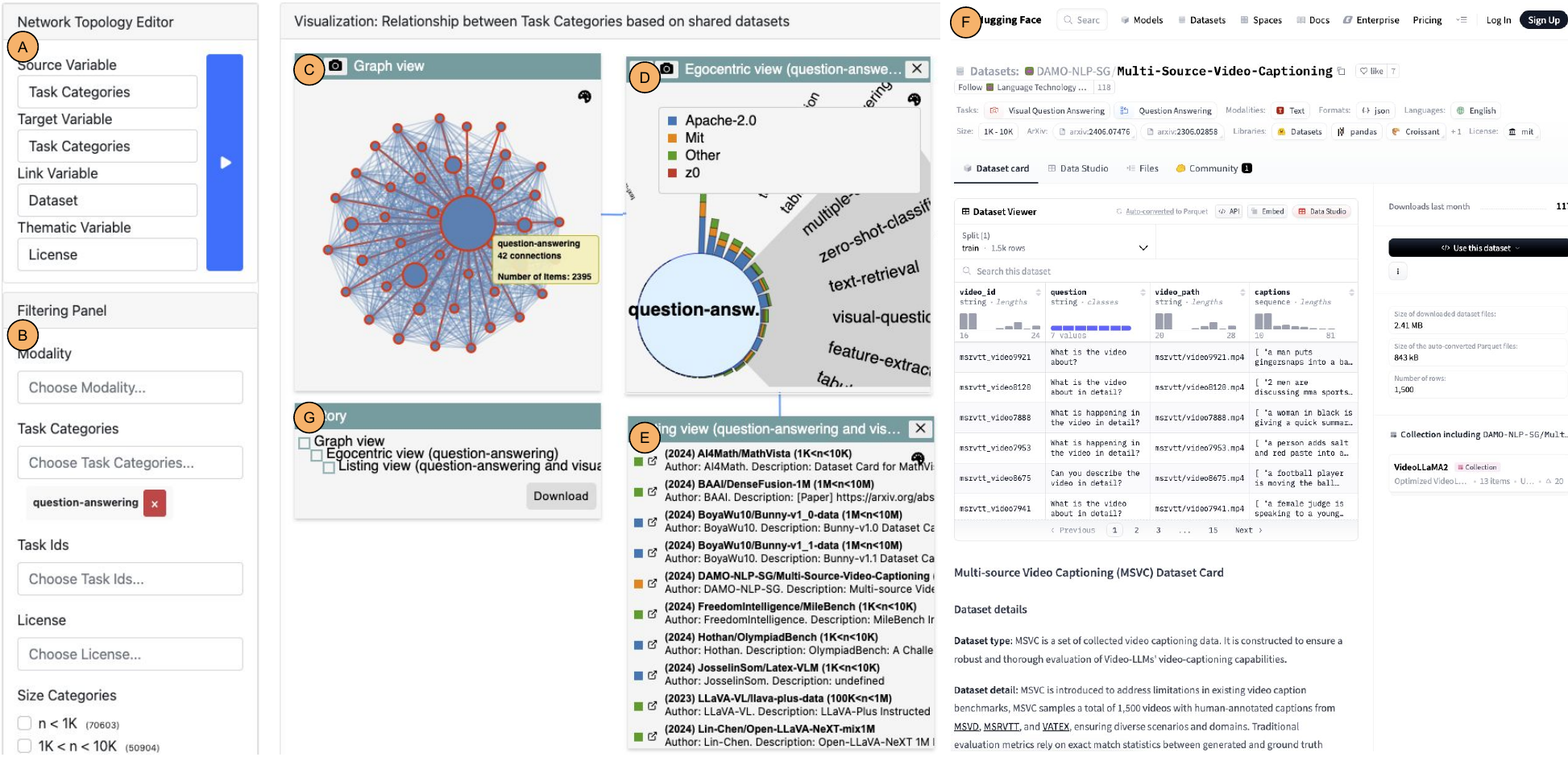}
    \caption{Overview of Use Case Scenario~\ref{ssec:scenario2}. (A) Task categories are represented as nodes, linked by datasets characterized by their distribution license. (B) Metadata is filtered to retain datasets covering at least the question-answering task. (C) The Graph View displays the network, with node size indicating the number of supporting datasets. (D) The Egocentric View enables focused, pairwise exploration of task relationships. (E) The Listing View presents datasets supporting both QA and visual-question-answering. (F) Direct dataset inspection at the source.}
    \label{fig:scenario2}
\end{figure*}

To explore additional QA-related tasks for extending her model, Alice returns to the network topology editor and sets \texttt{task categories} as nodes (i.e. both source and target variables) and \texttt{dataset} as links, with \texttt{license} as the thematic attribute (Fig.~\ref{fig:scenario2}A). She keeps the \texttt{question-answering} filter for \emph{Task Categories}, and deletes the size-based filter (Fig.~\ref{fig:scenario2}B). This configuration allows her to visualize shared QA-related tasks based on the datasets that support them, regardless of their size.

The generated network represents tasks as nodes, connected when at least one dataset supports them. Since the data is filtered by the QA task, this task appears as the largest node, with its size reflecting the number of supporting datasets. By hovering over the QA node, Alice discovers that these datasets also support 42 other distinct tasks (Fig.~\ref{fig:scenario2}C). She explores this subset further by right-clicking on the QA node and opening the \textit{Egocentric View} (Fig.~\ref{fig:scenario2}D).  

The \textit{Egocentric View} places the selected node (\texttt{question-answering}) at the center, displaying its connected nodes along a peripheral arc and highlighting relationships in a pairwise manner. Between each pair of nodes, a bar represents the number of shared links (i.e., datasets), with its length indicating the strength of the connection and its color segments reflecting the selected thematic attribute (e.g., distribution license).  

Alice is particularly interested in datasets that support both QA and \texttt{visual-question-answering} (VQA). To investigate further, she right-clicks on the bar linking these two tasks and opens the \textit{Listing View}, which displays the datasets supporting both QA and VQA (Fig.~\ref{fig:scenario2}E). Since she exclusively works with MIT-licensed datasets, she visually scans the list for a dataset marked with the corresponding color (5th in the list). Finally, by clicking the external link icon \faExternalLink, she accesses the dataset’s page on the source platform (e.g., Hugging Face) to inspect and download it (Fig.~\ref{fig:scenario2}F).


\subsection{Uncovering Dataset Relationships Through Shared Pretrained Models}
\label{ssec:scenario3}

\begin{figure*}[!ht]
	\centering
	\includegraphics[width=.9\linewidth]{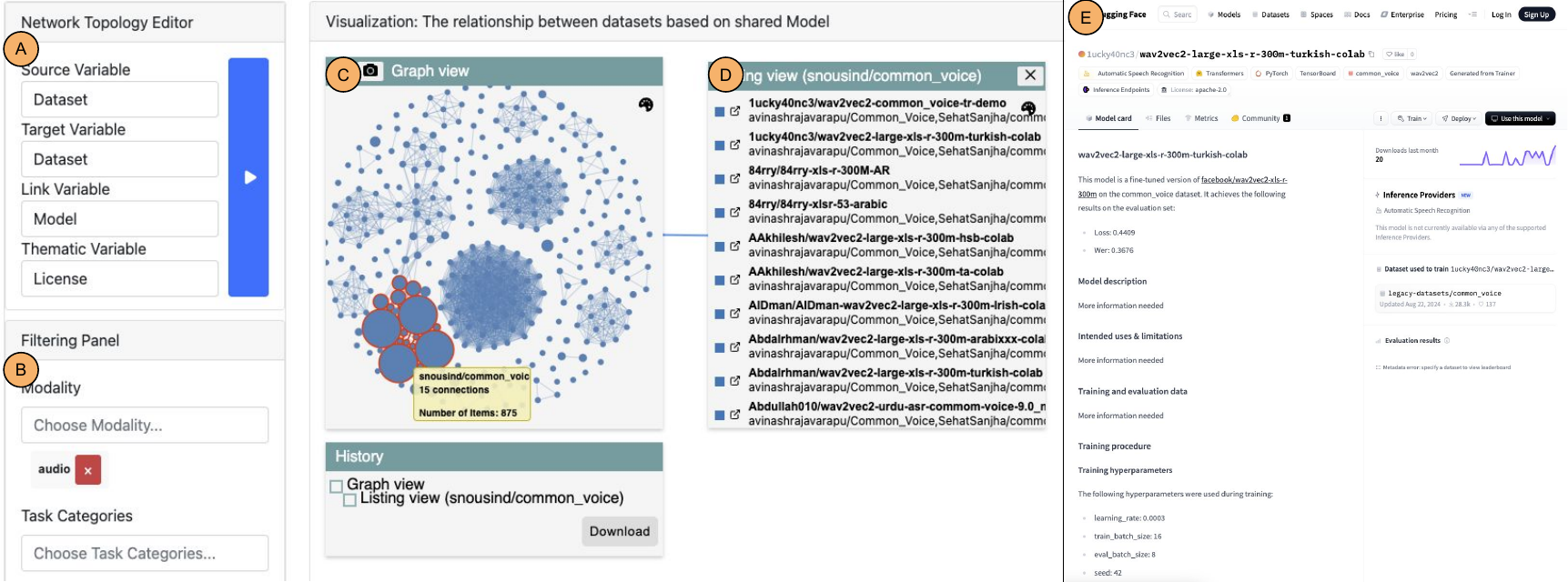}
	\caption{Use Case Scenario \ref{ssec:scenario2}. (A) Datasets are nodes, models are links. (B) Filtering by \emph{audio} modality. (C) Node size shows the number of models trained by each dataset. (D) Models trained with \texttt{snousind/common\_voice}. (E) Direct model inspection at the source.}
	\label{fig:scenario2}
\end{figure*}

Bob aims to train his own models for audio-based applications by identifying benchmark datasets based on their training usage across multiple pretrained models. He begins his exploration by defining a network topology that aligns with his objective. He sets \texttt{datasets} as (source and target) nodes and \texttt{models} as links (Fig.\ref{fig:scenario2}A). 
To refine his analysis, Bob filters the data to include only datasets categorized under \texttt{audio} (Fig.\ref{fig:scenario2}B).

Bob launches the visualization by clicking \emph{play}, generating a graph that connects datasets based on shared models. He initially observes many datasets with no connections, indicating that their associated models were not trained using any other datasets. By observing the node sizes, Bob identifies promising datasets that have been used to train over 800 models—for example, \texttt{snousind/common\_voice} (Fig.~\ref{fig:scenario2}C), which has contributed to training 875 models. He inspects this dataset using the \emph{Listing View} (Fig.~\ref{fig:scenario2}D), which reveals the list of models trained using this dataset.  
%
By clicking the external link icon \faExternalLink next to one of the models, Bob can access detailed information directly on Hugging Face, allowing him to verify metadata and gather additional insights (Fig.\ref{fig:scenario2}E).

\section{Formative Evaluation}

We conducted a formative evaluation with six data practitioners specializing in NLP, Knowledge Engineering, and Software Engineering. Their common challenge was finding suitable datasets for research. Participants included those searching for ML datasets, open datasets, and coding datasets. Four had over seven years of experience on their research domain, while the others had one to three years. The evaluation consisted of three phases (see Appendix~\ref{annex:questionnaire}): (i) a questionnaire on participants' needs, habits, and challenges in dataset search, (ii) hands-on use of Datalens to answer aforementioned scenario-based questions, such as identifying datasets supporting both text and tabular modalities, and (iii) a final questionnaire assessing usability, user satisfaction, and improvement suggestions.

\textbf{User Needs, Habits and Challenges:} Participants primarily search for datasets in repositories like Hugging Face and Kaggle, academic papers, and code repositories (e.g., GitHub). Key criteria for selection include domain, modality, license, size, supported tasks, language, and models, while aspects like related publications, source, popularity, and publication date are considered less important. Data quality and curation were also emphasized as important aspects when searching datasets. Participants commonly face difficulties in finding relevant datasets due to issues like label mismatches, noisy data, inconsistent definitions, partial availability, and dataset discrepancies. Assessing dataset quality often requires extensive manual inspection, and data cleansing can be time-consuming. Most participants were unfamiliar with using visualization tools in dataset searches, with mixed opinions on its potential usefulness. When asked about desired features in an exploration tool, participants expressed a need for standardized, comprehensive listings of dataset characteristics, precise search capabilities, dataset previews, metadata exploration, and insights into usage by others. They would also like to have dataset comparison, similarity search, and visualizations to identify gaps or patterns. Linking datasets to related papers, code, and models was considered valuable for context and validation. Datalens addresses some of these needs by revealing similarities between datasets and supporting comparisons through visualizations.

\textbf{Datalens' Usability and Relevance:} We used the well-known SUS questionnaire~\cite{brooke_sus_1996} to assess the usability of our tool, which received an average score of 61.66/100. The positive aspects highlighted in the feedback included good feature integration and participants' willingness to use the tool regularly. However, improvements are needed to enhance accessibility and ease of use. Participants noted that the chosen visualization tool, MGExplorer, while useful, may be too complex for its intended purpose. Additionally, they felt a need to learn many aspects of the system before use, which increased the learning curve. Despite these challenges, participants were generally neutral about the difficulty of completing the tasks, with one finding the third task particularly difficult. All participants agreed on the usefulness of the tool's features (visualization, network editor, and faceted search), with the network topology editor being especially helpful for most tasks.

\textbf{User Appreciation and Improvements:} Participants appreciated Datalens for its intuitive graph view, smooth transitions, customizable multi-window organization, and ability to explore the search space. They valued its responsiveness, powerful filters, and the network topology editor and egocentric view. The tool effectively visualizes links between datasets, models, and tasks, with cardinality estimation. Key strengths included overview-zoom-filter support, a graph-based overview, and direct links to datasets/models. Participants also liked configuring multiple search variables and text-based graph searches. However, they raised concerns about unclear parameters, lack of linking-and-brushing, and the small default window. Labels were seen as opaque, with a request for icons. For some tasks (i.e. Scenario~\ref{ssec:scenario2}), graph visualization was distracting, with a preference for list-based views. Dense graphs and zooming were unhelpful for focusing on nodes, and participants preferred single clicks. Memorizing names and interface complexity were issues. Two separate search views were confusing, and zooming needed improvement. Participants suggested improvements like a page with more details, enabling the AND operator in the browser, larger graph view, and enhanced zooming to improve node selection.

\section{Discussion and Future Work}

In this paper, we presented \emph{Datalens}, a visualization approach to support dataset search leveraging multi-faceted search and network-based visualizations. We address dataset search by revealing hidden relationships through graph topologies, which expose patterns, structures, and interconnections. Beyond exact matches, our approach highlights links and overlaps between communities, helping users identify reusable and transferable datasets. We demonstrated the effectiveness of \emph{Datalens} through use case scenarios and a formative evaluation with data practitioners using HuggingFace dataset catalog, as a case study. 

While graph topologies reveal hidden links between datasets, they alone may not suffice for dataset search. As shown in our use case scenarios, multiple data dimensions and attributes require viewing the data from different perspectives. Thus, we provide multiple visualization techniques that present the relationships in complementary ways, while allowing users to refine their search, transitioning from one view to another. 
Our formative evaluation has confirmed the promising aspect of visualization to support dataset search, particularly, through the network visualization but also via an egocentric visualization showing detailed pairwise relationships between datasets. Participants appreciate the use of chained views to support the discovery process, but noticed that the tool (i.e. MGExplorer) might be too complex to the intended goal. Furthermore, the tool suffers from scalability issues, 
as the network can become rapidly cluttered when there are too many nodes and links to be represented. Furthermore, while the visualizations are useful, starting with a network was not always seen as necessary during the evaluation. Thus, future work includes the investigation of aggregation and interaction methods to tackle scalability issues, as well as studying the relevance of visualization techniques to support different dataset search tasks. 

We developed \emph{Datalens} in a manner that it adapts itself based on available metadata. Thus, while we currently applied our approach to datasets published on Hugging Face, one can easily extend and apply the tool to any other dataset repository, such as \emph{Kaggle}. In the spirit of open science, the source code is publicly accessible on GitHub\footnote{\url{https://github.com/amenin/datalens}}. As future work, to extend the coverage of \emph{Datalens} and support dataset search in other domains, such as software engineering, we intend to dynamically integrate other dataset repositories.

Metadata quality issue remains a challenge in dataset discovery. For example, in the Hugging Face catalog, labels are only partially standardized, allowing users to add custom descriptions, which leads to heterogeneous dataset profiling. Furthermore, there are multiple attributes that do not have associated values, which hinders dataset discovery as we cannot classify it easily (e.g., over 57.7k datasets do not have an associated modality). To address this issue and enhance interoperability, future work includes integrating domain-specific metadata standards with controlled vocabularies, as recommended by the RDA Data Maturity Model\footnote{\url{https://www.rd-alliance.org/}}. We also intend to investigate the usage of semantic web technologies to model and represent dataset metadata in a inherently interoperable way.  Additionally, we aim to enrich metadata by answering key questions (e.g., What is the dataset for? Who is it for? Why and when is it used?), focusing on data seekers' needs. 





\bibliography{references}

\newpage
\appendix
\onecolumn

\footnotesize
\setlength{\topsep}{0pt}        
\setlength{\itemsep}{0pt}       
\setlength{\parsep}{0pt}        
\setlength{\partopsep}{0pt}     

\section{Questionnaire/Protocol used during the Formative Evaluation}
\label{annex:questionnaire}

\subsection{Terms and Conditions Agreement}

The goal of this user study is to assess the capabilities of DataLens, a web-based platform that integrates faceted search with advanced information visualization techniques, to facilitate the search and exploration of machine learning (ML) datasets. Specifically, we aim to explore the following:

Identifying user tasks and needs, with a focus on the types of queries and attributes essential for effectively searching and selecting ML datasets.

Evaluating the added value of leveraging both customizable network topologies and multi-perspective visualizations to effectively meet the needs of data practitioners.

Investigating the benefits of visual exploration of dataset networks through information visualization techniques. In this study, we use the visualization tool MGExplorer.

This study will last around 40 minutes and will be structured as follows. We will start with a few questions to learn about your profile, habits, and needs regarding the usage and exploration of ML datasets. Next, we will introduce you to the Datalens interface and its main features. Then, we will ask you to complete three exploratory tasks using either the Datalens interface or the HuggingFace search feature to explore datasets from the HuggingFace catalog. Lastly, we will collect your feedback and comments on the tool's capabilities.

Participation in this study is anonymous. The collected data will be used in this study only. You may withdraw at any time without providing a reason. If you choose to do so, your data will be immediately deleted and will not be used in this study.

Please, check the option below if you consent to participate in this study according to the conditions presented above.

\begin{itemize}
    \item[\RadioButton] I consent to participate in this study. 
\end{itemize}

\subsection{Pre-questionnaire}

\begin{enumerate}
    \item What is your main research area?
    \item How many years of experience do you have in your field?\\
    \RadioButton Less than 1 year\quad \RadioButton Between 1 and 3 years\quad \RadioButton Between 4 and 6 years\quad \RadioButton Over 7 years
    \item How often do you search for ML datasets? (5-point Likert scale)
    \item Where to you usually search for datasets?\\
    \CheckBox Dataset Repositories (e.g. Kaggle, HuggingFace) \CheckBox Academic Papers \CheckBox Code repositories (e.g. Github) \CheckBox Personal or professional networks
    \item How important are the following elements to you when searching for datasets? (5-point Likert scale)
    \begin{itemize}
        \item The domain or application area of the dataset.
        \item The size of the dataset (e.g., number of records, file size).
        \item The language(s) represented in the dataset.
        \item  The modality of the dataset (e.g., text, audio, image).
        \item Licensing information (e.g., open-source, commercial).
        \item The time of data collection.
        \item The source of dataset (e.g. institution, authors, etc.).
        \item The popularity of the dataset (e.g. likes, downloads).
        \item Tasks for which the dataset has been applied.
        \item Scientific publications citing or utilizing the dataset.
        \item Models trained using the dataset.
        \item Temporal trends in dataset usage.
    \end{itemize}

    \item Are there any other factors that you consider important when searching for datasets? Please specify.
    \item How often do you encounter problems when searching for datasets? (5-point Likert scale)
    \item What king of problems do you encounter? If applicable.
    \item How often do you use visualization tools to search datasets? (5-point Likert scale)
    \item How important do you consider visualization techniques to support dataset search? (5-point Likert scale)
    \item What features would you like to see in a dataset search and exploration tool?
\end{enumerate}

\subsection{Hands-on use of Datalens}
Go to \href{http://dataviz.i3s.unice.fr/datalens/explorer}{Datalens: http://dataviz.i3s.unice.fr/datalens/explorer}.

\subsubsection*{Task 1: Identifying Suitable Datasets for a QA Chatbot}

Your objective is to find datasets for developing a Question Answering (QA) model for a chatbot. Specifically, you want to explore available datasets and their modalities (e.g., text or audio) to define how users will interact with the chatbot. Follow the steps below:

\begin{enumerate}
    \item Configure the Network Topology as follows:
    \begin{itemize}
        \item \textbf{Source variable}: Dataset
        \item \textbf{Target variable}: Modality
        \item \textbf{Link variable}: Task Categories
        \item \textbf{Thematic attribute}: default (not applicable)
    \end{itemize}
    \item To keep a set of interesting datasets to explore, filter the metadata to consider only QA tasks and relatively large datasets. For that, select the following filters:
    \begin{itemize}
        \item \textbf{Task Categories}: question-answering
        \item \textbf{Size Categories}: $1M < n < 10M$
    \end{itemize}
    \item Click on the \textbf{play} (blue button) to launch the visualization.
\end{enumerate}

\textbf{Question:} Can you identify 2 datasets that you can use for both tabular and textual data?  

\subsubsection*{Task 2: Identifying Related Tasks to Expand your Model's Applications}

Still within the goal of developing a QA model for a chatbot application, your objective now is to investigate related or adjacent sub-tasks to expand your model's potential applications. Follow the steps below:

\begin{enumerate}
    \item Configure the Network Topology as follows:
    \begin{itemize}
        \item \textbf{Source variable}: Task Categories
        \item \textbf{Target variable}: Task Categories
        \item \textbf{Link variable}: Dataset
        \item \textbf{Thematic attribute}: License
    \end{itemize}
    \item To keep a set of interesting datasets to explore, filter the metadata to consider only QA tasks and expand the search to include all possible sizes of datasets. For that, select the following filter:
    \begin{itemize}
        \item \textbf{Task Categories}: question-answering
    \end{itemize}
    \item Click on the \textbf{play} (blue button) to launch the visualization.
\end{enumerate}

\textbf{Question:} Can you find 2 datasets that support both Question Answering (QA) and Visual Question Answering (VQA) and that are available under the MIT License? Share below the links to their descriptions on HuggingFace.

\subsubsection*{Task 3: Identifying Relevant Datasets based on the Models that use them}

In this task, your goal changes as you aim at improving your model selection process for audio-based applications. For that, you will try to identify the most popular datasets, based on their shared usage across models. Follow the steps below:

\begin{enumerate}

    \item Configure the Network Topology as follows:
    \begin{itemize}
        \item \textbf{Source variable}: Dataset
        \item \textbf{Target variable}: Dataset
        \item \textbf{Link variable}: Model
        \item \textbf{Thematic attribute}: default (not applicable)
    \end{itemize}
    \item To keep a set of interesting datasets to explore, filter the metadata to consider only datasets within the audio modality. For that, select the following filter:
    \begin{itemize}
        \item \textbf{Modality}: audio
    \end{itemize}
    \item Click on the \textbf{play} (blue button) to launch the visualization.
\end{enumerate}

\textbf{Question:} Can you identify one widely used dataset for training models (i.e. one associated to many models) and three models that use it? Provide below the links to the models' descriptions on HuggingFace.

\subsubsection*{Post-task questions (after each task)}

\begin{itemize}
    \item How do you evaluate the difficulty of this task? (5-point Likert scale)
    \item How useful did you find the following elements to solve this task? Network topology, filters, visualizations (5-point Likert scale)
    
\end{itemize}

\subsection{Post-questionnaire}

\begin{enumerate}
    \item SUS questionnaire \cite{brooke_sus_1996}
    \item Please provide 3 things that you \textbf{liked} about the experience or the tool.
    \item Please provide 3 things that you \textbf{disliked} about the experience or the tool.
    \item What enhancements or new features would you recommend to improve the tool's usefulness and user experience?
\end{enumerate}



\end{document}